%% file: main.tex
\documentclass[12pt]{iopart}

\usepackage{upgreek} 
\usepackage{graphicx}
\usepackage{float}
\graphicspath{ {./Figures/} }
\newcommand\COtwo{CO$_2$} 
\newcommand\Otwominus{O$_2^-$} 
\newcommand\Omin{O$^-$} 
\newcommand\COthree{CO$_3^-$} 
\newcommand\tlv{$t_{\rm LV}$} 
\newcommand\vlv{$V_{\rm LV}$} 
\newcommand\tlveq{t_{\rm LV}} 
\newcommand\vlveq{V_{\rm LV}} 
\newcommand\vhveq{V_{\rm HV}} 
\newcommand\tinc{$t_{\rm inc}$} 
\newcommand\tmed{$\tilde{t}_{\rm inc}$}
\newcommand\mug{$\upmu$} 
\newcommand\cotwomin{CO$_2$$^-$} 
\newcommand\elec{e$^-$} 
\newcommand\layer{charged species patch} 

\begin{document}

\title[Discharge inception in \COtwo{} ]{Investigating \COtwo{} streamer inception in repetitive pulsed discharges }

\author{S. Mirpour and S. Nijdam}

\address{Department of Applied Physics, Eindhoven University of Technology, PO Box 513, 5600 MB Eindhoven, The Netherlands}
\eads{s.mirpour@tue.nl}
\vspace{10pt}

\begin{abstract}
\input{abstract-CO2}   
\end{abstract}

\noindent{\it submitted to\/} Plasma Sources Science and Technology, Version \today
\maketitle
%
%
%
%
%

\input{body-CO2}
\section*{Acknowledgement}
We would like to thank Andy Martinez, Prof.Dr. Ute Ebert, and Dr. Jannis Teunissen from Centrum Wiskunde \& Informatica (CWI), who provided insight and expertise that greatly assisted the research. This project has received funding from the European Union's Horizon 2020 research and innovation program under the Marie Sklodowska-Curie grant agreement 722337.
\section*{References}
\bibliography{main}
\end{document}

%% file: abstract-CO2.tex
In this study, we investigate the responsible species and processes involved in repetitive pulsed streamer inception in \COtwo{}. We applied a 10\,kV high-voltage pulse with a repetition frequency of 10\,Hz and pulse width of 1\,ms to a pin electrode which is placed 160\,mm apart from the grounded plane electrode. We measured the inception times (delay between the rising edge of the high-voltage pulse) by a photo-multiplier tube for 600 high voltage cycles. We observed one peak in the histogram of inception times with a median of 1.2\,\mug{}s. To identify the source of this peak, we applied a negative or positive LV pulse before the main HV pulse to manipulate the leftover space charges. Three different phenomena are observed: 1) drift, 2) neutralization, and 3) ionization in the LV pulse. At low LV amplitude and pulse width, the peak starts to drift toward the faster and slower inception times under a positive and negative LV pulse, respectively. However, under the same LV pulse configuration for positive and negative LV pulse, the observed shift in inception times is not the same. We present a hypothesis to explain this asymmetry based on the difference of the detachment processes between air and \COtwo{}.   

%% file: body-CO2.tex
\section{Introduction}

Streamers are widely explored in different gases because of their importance in various fields such as high-voltage engineering and atmospheric electricity \cite{nijdam2020physics,ebert2008streamers,ebert2006multiscale,nam2009empirical,kolb2008streamers,sun2016formation,akiyama2000streamer,Bonaventura_2011}. Recently, \COtwo{} gas has been considered as a possible candidate for high-voltage switchgear, since the gas currently used for insulation and switching, SF$_6$, is an extremely strong greenhouse gas \cite{seeger2017recent,kieffel2015sf}. Also, there is strong evidence that lightning might exist in the Martian atmosphere which mostly consists of \COtwo{} \cite{ruf2009emission,melnik1998electrostatic}. Hence, it is important to understand the physics behind streamer inception and development processes in this gas. Streamer inception voltage and time have been mostly studied in air \cite{mirpour2020distribution,pedersen2009streamer,briels_time_2008,teunissen_3d_2016}. There are not many studies of discharge inception in \COtwo{} (see below), and the current understanding of the streamer inception process is very limited. This study investigates the streamer inception process in \COtwo{} in more detail.

By applying positive high-voltage pulses to an electrode, electrons start to drift opposite to the electric field. When such an electron reaches an area  where the ionization rate is higher than the attachment rate, it can start to replicate, although this is of course a stochastic process. The electron density then grows until the formed space-charge becomes large enough to transform into a streamer. The number of electrons required for this transition can be derived from the so-called Meek criterion \cite{montijn2006diffusion,Meek1940}.
In our previous work \cite{mirpour2020distribution}, we hypothesized that in synthetic air, the very first electron to trigger the streamer inception process can come from three different sources: free electrons, \Otwominus{} detachment, or from reactions of neutral species. We showed that, in repetitive discharges, the leftover charges from the streamers likely build up an in-homogeneous distribution of ions in front of the anode, which produces a peak in the streamer inception distribution times histogram. By adding a low-voltage (LV) pre-pulse before the main high-voltage (HV) pulse we could manipulate and drift this in-homogeneous ion cloud towards or away from the anode, depending on the LV pulse polarity. Thus, the ions will be at a new position at the beginning of the HV pulse which results in a shift of the observed peak in the time distribution histogram. From these shifts, we can derive the mobility of ions and identify the source. 
Photoionization is another well-known source of initial seed electrons in air. However, no efficient photoionization mechanism is known for \COtwo{} due to the rapid absorption of photons in this gas \cite{bagheri2020simulation}. Therefore, in \COtwo{}, detachment of electrons from negative ions is the most probable source of initial free electrons.   

The statistical time lag for \COtwo{} discharges has been studied by Seeger et al.~\cite{seeger2016streamer}. They showed that, at pressures higher than ambient pressures, by increasing the applied voltage, the discharge inception time lag decreases and reaches as low as 100\,ns. They found a fairly good agreement between the theoretical prediction of streamer inception fields (considering the ionization integral as 13) and actual observations. However, that study offers little in the way of examining which species are playing an important role in inception.

Wang et al. \cite{wang2016effective} investigated the critical breakdown field for \COtwo{} discharges at elevated temperatures. They showed that at room temperature the reduced breakdown electric field in \COtwo{} is about 86\,Td. Increasing the temperature leads to \COtwo{} dissociation, which leads to more collisions between CO and O$^-$ creating electrons through associative detachment. This will influence the effective ionization rate, contributing to a significant drop in the critical breakdown field. At room temperature this mechanism is insignificant due to the low dissocation degree. Furthermore, \Omin{} is turned into a relatively stable ion, \COthree{}, by the cluster stabilization process. Hence, the main electron source and sink are electron impact ionization and dissociative attachment to \COtwo{} through \elec{} + \COtwo{} $\rightarrow$ CO + \Omin{}.

This study aims to clarify the role of different species in the streamer inception process in \COtwo{} by adding a pre-pulse to manipulate the charge distribution before applying the main HV pulse. The effects on these on inception time histograms can give great insights in the relevant processes. This paper is organized as follow: In section 2 we describe the method, section 3 discusses the results and gives an hypothesis based on the different LV pulse configurations. Finally, in section 4 we summarize the paper and provide the conclusion.

\section{Set-up and Methods}
\subsection{Experimental conditions}
\begin{figure}[ht!]
    \centering
    \includegraphics[scale=0.8]{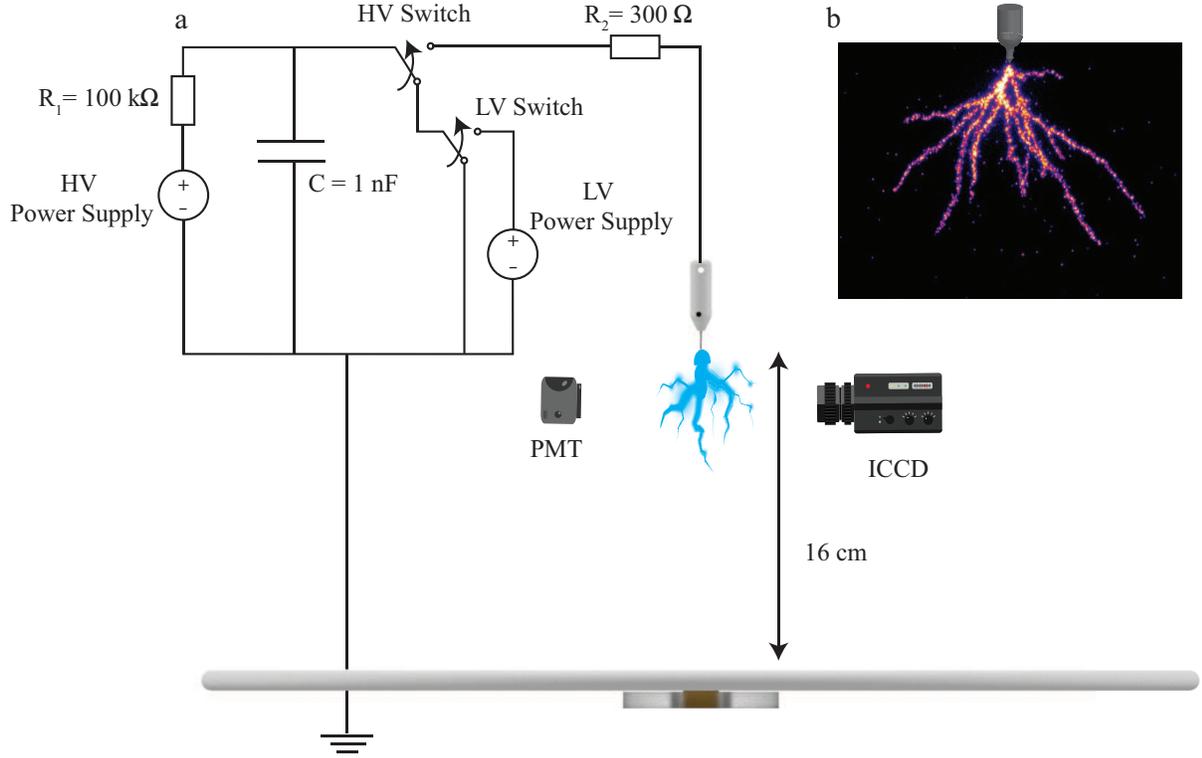}
    \caption{a) Schematic view of the experimental setup with HV power supply connected to the anode (not to scale). b) ICCD image of the \COtwo{} discharge at 300\,mbar with HV amplitude of 10\,kV and repetition frequency of 10\,Hz.}
    \label{fig:setup_co2}
\end{figure}

All experiments in this study were performed in a point-to-plan geometry (figure~\ref{fig:setup_co2}) in which the powered electrode, (anode
with a tip radius of about 100\,$\upmu$m), is separated 160\,mm from the grounded electrode. The anode is powered by a circuit comprising an HV solid state switch (Behlke, HTS 651-15-SiC-GSM) and a 1\,nf capacitor. With this configuration, we can produce repetitive pulsed waveforms with an amplitude of 10\,kV, pulse widths of 1\,ms, rise times of 40\,ns, and repetition rates of 10\,Hz. The reason we chose this repetition rate and pulse duration is to have only one discharge within the duration of an HV pulse. We did not observe any dependency of streamer inception on pulse duration at this frequency.  The background pressure was $10^{-6}$\,mbar, and for each experiment the vessel was filled to 300\,mbar with \COtwo{} (purity level of 99.9999\,\%).\\ 

\begin{figure}[ht!]
    \centering
    \includegraphics[scale=0.8]{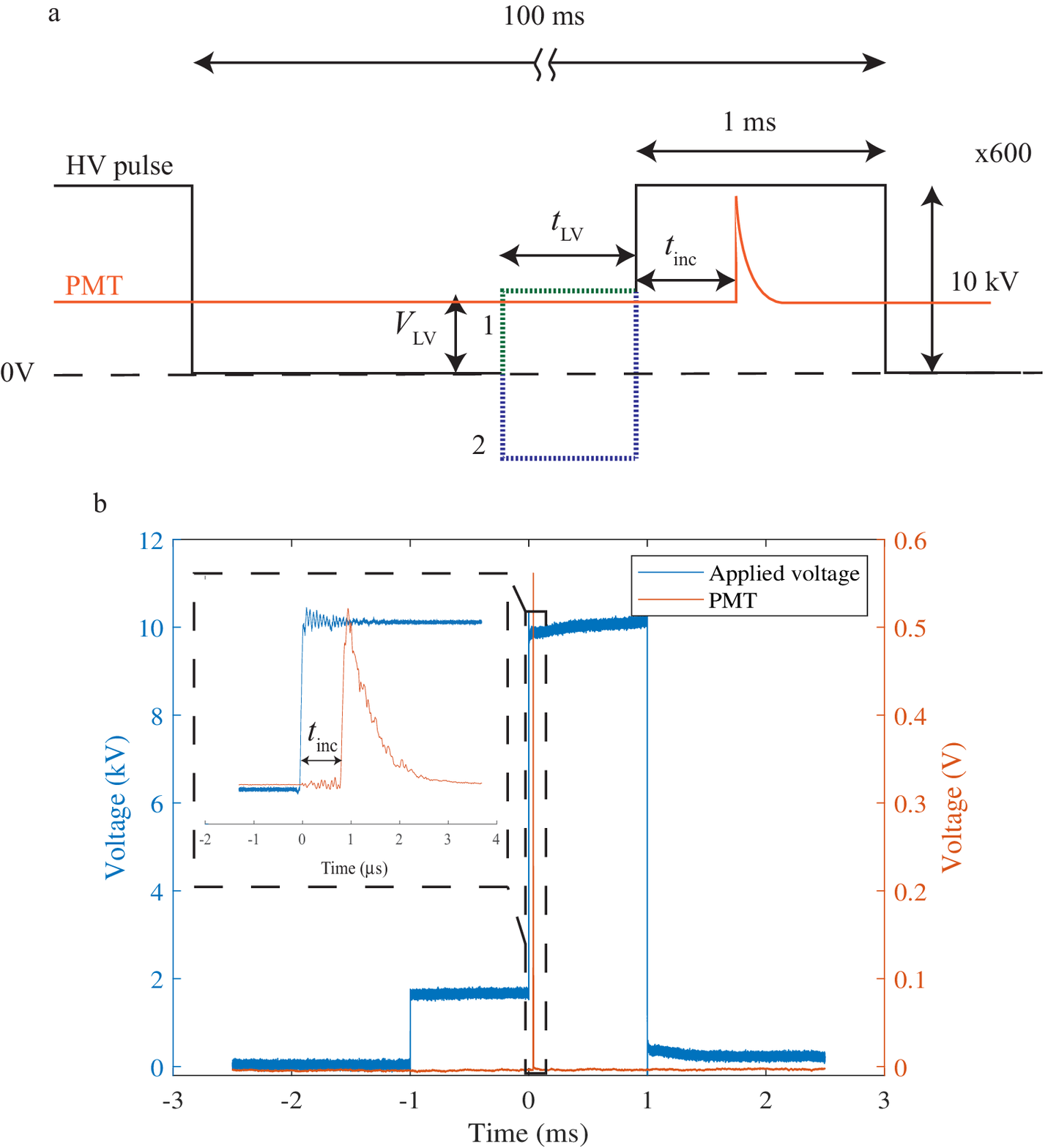}
    \caption{a) Scheme of applied HV waveform with 1. positive and 2. negative attached LV pulse. b) Typical applied HV and LV voltage with PMT waveform output.}
    \label{fig:waveforms}
\end{figure}

To investigate the influence of residual charged species on streamer inception, an LV pre-pulse was applied before each HV pulse. To apply the LV pre-pulse, the negative input of the HV pulser was connected to a second pulser, comprised of another solid-state HV switch (Behlke, HTS 181-01-HB-C) and a 1\,nF capacitor. This could produce a pre-pulse which attaches to the main HV pulse and has an amplitude \vlv{} between 0 and 8\,kV and a pulse width \tlv{} between 0 and 100\,ms (c.f. figure~\ref{fig:waveforms}).

\subsection{Measuring inception time \tinc{}}
To measure the streamer inception time (\tinc{}), we used a photo-multiplier tube (PMT,H10720-1 10 Hamamatsu) placed behind a window of the vessel. The response time of the PMT is less than 2\,ns. The output signal was transferred to a 12-bit HD 6104 Teledyne Lecroy oscilloscope with a sample rate of 1\,GS/s. We repeated this measurement for 600 HV cycles per setting. Note that we consistently observed only one discharge inception event per 1\,ms HV pulse. 
For each cycle, \tinc{} was determined as the temporal delay between the HV pulse reaching 10\,\% of its maximum and the PMT waveform reaching 10\,\% of its maximum (see figure~\ref{fig:waveforms}). The total error in \tinc{} is estimated at less than 10\,ns. Streamer inception is here defined as an HV pulse in which the PMT signal shows a peak three times above the average background noise.

From the measured \tinc{} values, a histogram is made using a logarithmic binning function, MATLAB function:\texttt{logspace(log10(0.01), log10(max(data)), 700)}, which divides the data into 700 bins starting from 10\,ns. The acquisition window was 1\,ms for all experiments.
In the histogram figures, we indicate the inception probability ($Prob$), which is the number of pulses for which streamer inception was detected, divided by the total number of HV pulses (600). An intensified CCD (ICCD, Stanford Computer Optics 4QuickE) with a nanosecond time gate and a Nikkor UV 105\,mm lens f/4.5 was used to image the streamers. The streamers in figure~\ref{fig:setup_co2}.b are rendered in a false-colour scale for clarity. We found that in each case that a peak was detected by the PMT, the ICCD image always showed a developed streamer.

\section{Results and discussion}
\subsection{Baseline experiment and general phenomenology}
\begin{figure}[ht!]
    \centering
    \includegraphics[scale=0.8]{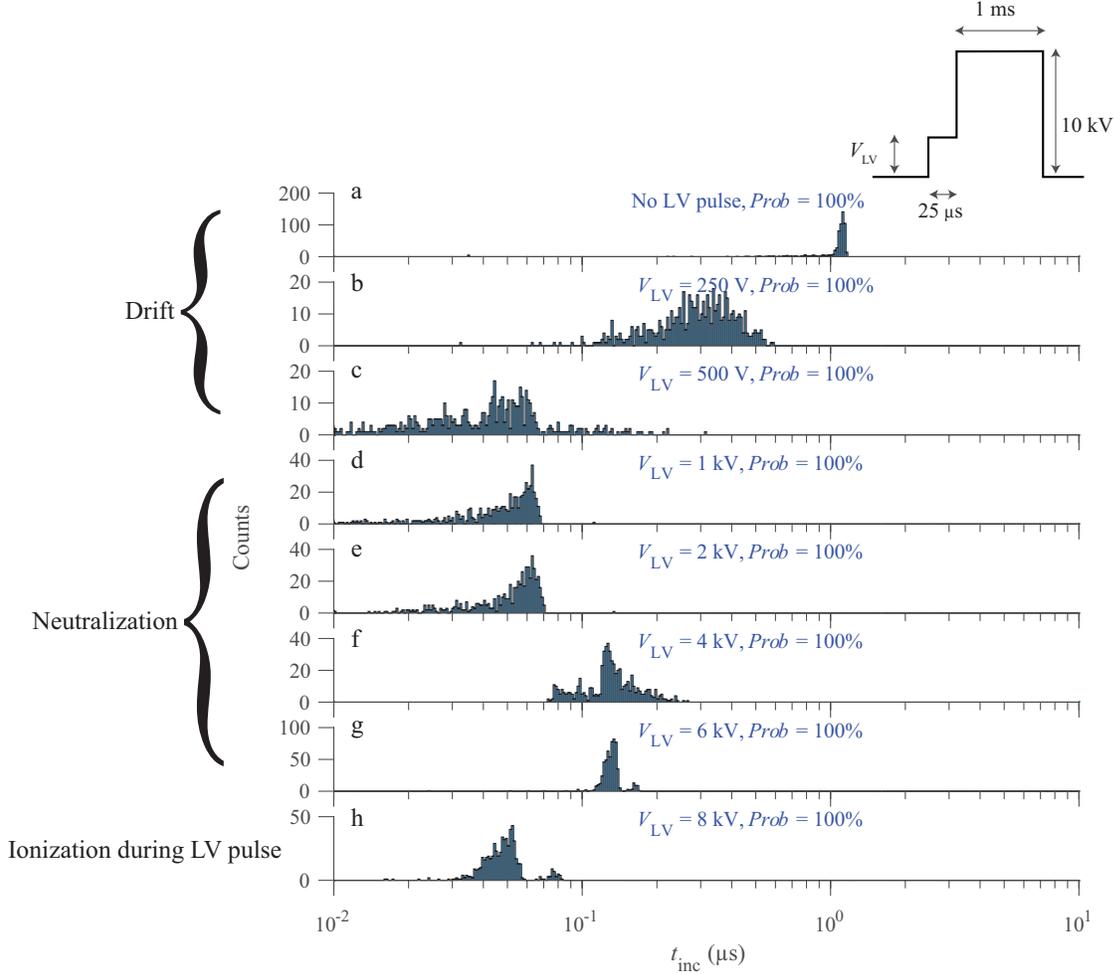}
    \caption{Histograms of discharge inception time \tinc{} for 600 discharges produced a) for no LV pulse and b-h) by applying a 25\,\mug{}s duration LV pulse with the indicated voltage before a 10\,kV HV pulse of 1\,ms with a repetition frequency of 10\,Hz. Inception probabilities are 100\,\% for each case, as indicated.}
    \label{fig:general_phen}
\end{figure}

To gain an overview of phenomenology, we applied an HV pulse (10\,kV amplitude with a duration of 1\,ms) and an LV pre-pulse (with a duration of 25\,$\upmu$s and varying \vlv{}) to the electrode. We repeated this 600 times at a repetition frequency of 10\,Hz. Figure~\ref{fig:general_phen} shows the histogram of \tinc{} on a logarithmic scale where the size of the time bins is scaled as log\,$t$. Figure~\ref{fig:general_phen}.a shows the baseline experiment with no LV pulse where \tinc{} forms a single peak around 1-2\,$\upmu$s. Application of a positive LV pulse before the main HV pulse led to three observations. For \vlv{} up to 500\,V (Drift, figure~\ref{fig:general_phen}.a-c) the \tinc{} peak started to shift to lower values, below 70\,ns. Increasing \vlv{} more (Neutralization, figure~\ref{fig:general_phen}.d-g) caused the peak to shift to higher values around 150\,ns. Finally, an LV pulse of 8\,kV (Ionization during LV pulse, figure~\ref{fig:general_phen}.h) again shifted the peak to 50\,ns. Note that this phenomenon depends on an interplay between \vlv{} and \tlv{} and therefore we introduce $S$= $\tlveq{}\cdot\vlveq{}$ as the main parameter in the phenomenology. Below, each of the observed phenomena is elaborated in a separate section.

\subsection{Drift due to applying an LV pulse}
\begin{figure}[h!]
    \centering
    \includegraphics[scale=0.8]{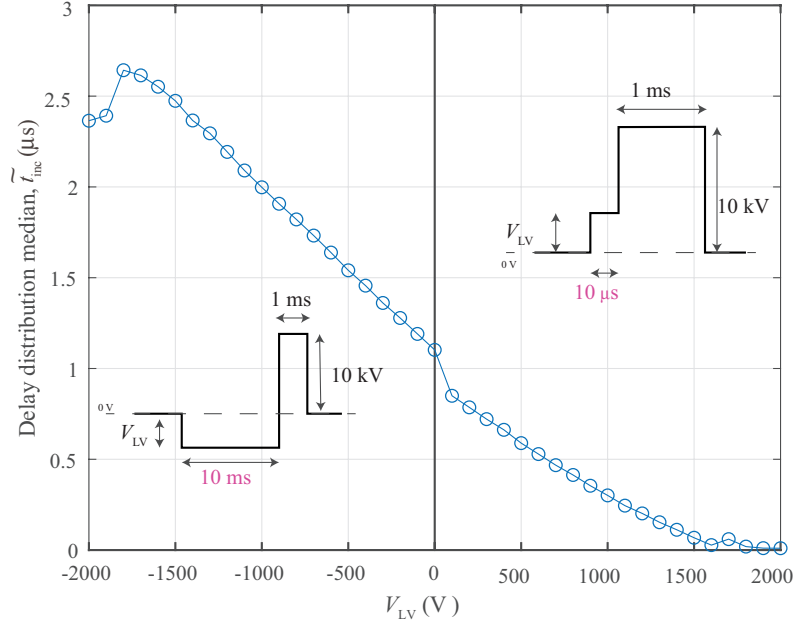}
    \caption{Median of the delay histograms, \tmed{}, produced by applying left) negative and right) positive LV pulses with different \vlv{}. For the negative and positive LV pulse, \tlv{} is fixed and set to be 10\,ms and 10\,$\upmu$s, respectively.  }
    \label{fig:media_volt}
\end{figure}

Since we mostly observe one peak in the \tinc{} histogram, the median values (\tmed{}) were calculated and are plotted against \vlv{} in figure~\ref{fig:media_volt}). To test the role of leftover charges on the inception, we applied both positive and negative LV pulses immediately before the main HV pulse. Figure~\ref{fig:media_volt} shows the effects of these LV pulses on \tmed{}, where for negative the LV pulses, a pulse width \tlv{} of 10\,ms was used, while for the positive LV pulses this was 10\,$\upmu$s, a factor 1000 lower. In the electric field imposed by a positive LV pulse, the negatively charged species will drift towards the electrode and thereby initiate discharges quicker after the HV pulse is applied. The results show that for every 100\,V of applied positive LV pulse, we observed a 94\,ns shift in \tmed{}. 

We would expect to have the same drift magnitude in opposite direction for a negative LV pulse. Interestingly however, a negative LV pulse with a 10\,\mug{}s pulse width leads to a negligible shift in \tmed{}. As can be seen in Figure~\ref{fig:media_volt}, to have a 90\,ns shift in \tmed{} we need to apply a 10\,ms, 100\,V negative LV pulse instead of a 10\,\mug{}s, 100\,V pulse. Note that \tmed{} is not fully linear with \vlv{} so the mentioned numbers are averages.    

\begin{figure}[h]
    \centering
    \includegraphics[scale=0.8]{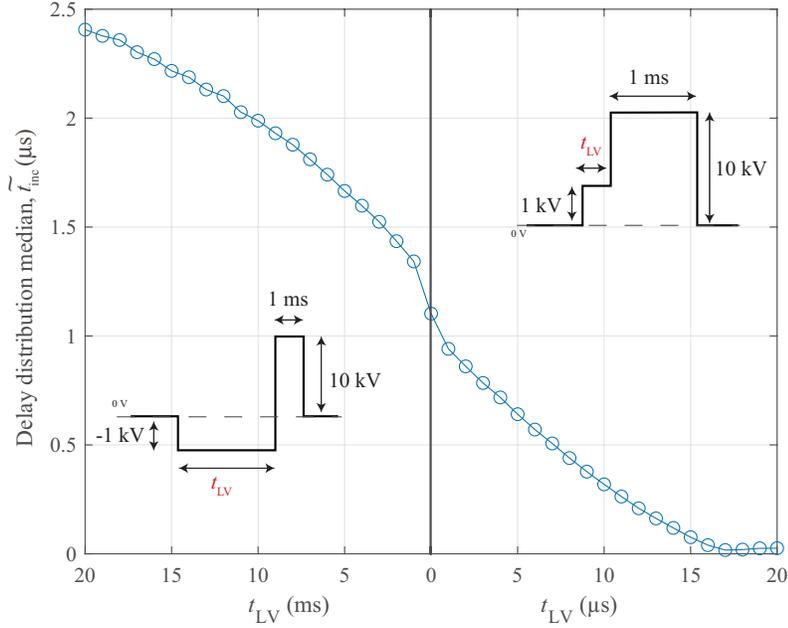}
    \caption{Median of the delay histograms, \tmed{}, produced by applying left) negative and right) positive LV pulses with different \tlv{}. For the negative and positive LV pulse, \vlv{} is fixed and set to be -1\,kV and 1\,kV, respectively. Note the difference in time units between left-hand and right-hand side on the horizontal axis.}
    \label{fig:media_time}
\end{figure}

To investigate the effect of the duration of the LV pulse, we fixed \vlv{} to 1\,kV and -1\,kV and varied \tlv{}. Figure~\ref{fig:media_time} shows the \tmed{} shift when applying positive and negative LV pre-pulses. For negative LV pre-pulses, a 1\,ms increase in \tlv{} leads to a 86\,ns decrease in \tmed{}. Similar to what we observed in the previous experiment \tmed{} the reverse effect for positive pre-pulses is also present but again roughly a thousand times stronger; to have a 92\,ns shift in \tmed{} we need to apply a 1\,kV positive pre-pulse with $\tlveq{}=1$\,\mug{}s instead of 1\,ms.
Note that in our previous experiments in air \cite{mirpour2020distribution} the effects of positive and negative pre-pulses were nearly equal, but of course opposite in direction.
Below, we give a hypothesis to explain this large asymmetry of three orders of magnitude between the effects of positive and negative pre-pulses in \COtwo{}. 

\subsubsection{Hypothesis}
In order to explain the surprising asymmetry between the effects of positive and negative pre-pulses in \COtwo{} (and not in air), we hypothesize that this is caused by the detachment process in \COtwo{}.

During an HV pulse, energetic electrons near the tip cause \COtwo{} dissociation by direct electron impact. The dissociative attachment reaction rate increases in higher electric fields and causes the production of CO and \Omin{} near the electrode \cite{hokazono1987theoretical}. Thus, in this specific region, both these species are available and recombination of \Omin{} and CO can liberate an electron through  associative detachment at any time during and between voltage pulses. During an HV pulse, such an electron can enter the ionization region and initiate a discharge. If there is a "\layer{}": a volume containing positive ions, negative ions and free electrons, located some distance from the tip, like we also concluded in \cite{mirpour2020distribution}, then the electrons from this region can again explain the peaked distribution of the baseline experiment. Note that from our results we cannot determine the shape of the \layer{}. It could be a thin layer, a localized patch, or a large region extending away from the electrode tip. We only know that it has a quite sharp boundary towards the electrode.

When applying a positive LV pulse, all negative species in the \layer{} will drift toward the tip. Since ions are much heavier than electrons, electron drift is more prominent and therefore the electron mobility will determine the effect of the LV pulse on inception behaviour.
A negative LV pulse with the same $S$ value will also push away the electrons from their initial position for roughly the same distance as the positive LV pulse. However, the much less mobile ions can still produce electrons by associative detachment and remain much closer to the tip than the drifted electrons due to their lower mobility. During the main HV pulse inception times are determined by the closest electrons and not the far-away ones. Therefore, for positive LV pulses, electron drift dominates the effect, while for negative LV pulses, negative ion drift does so.

\begin{figure}[ht!]
    \centering
    \includegraphics[scale=0.8]{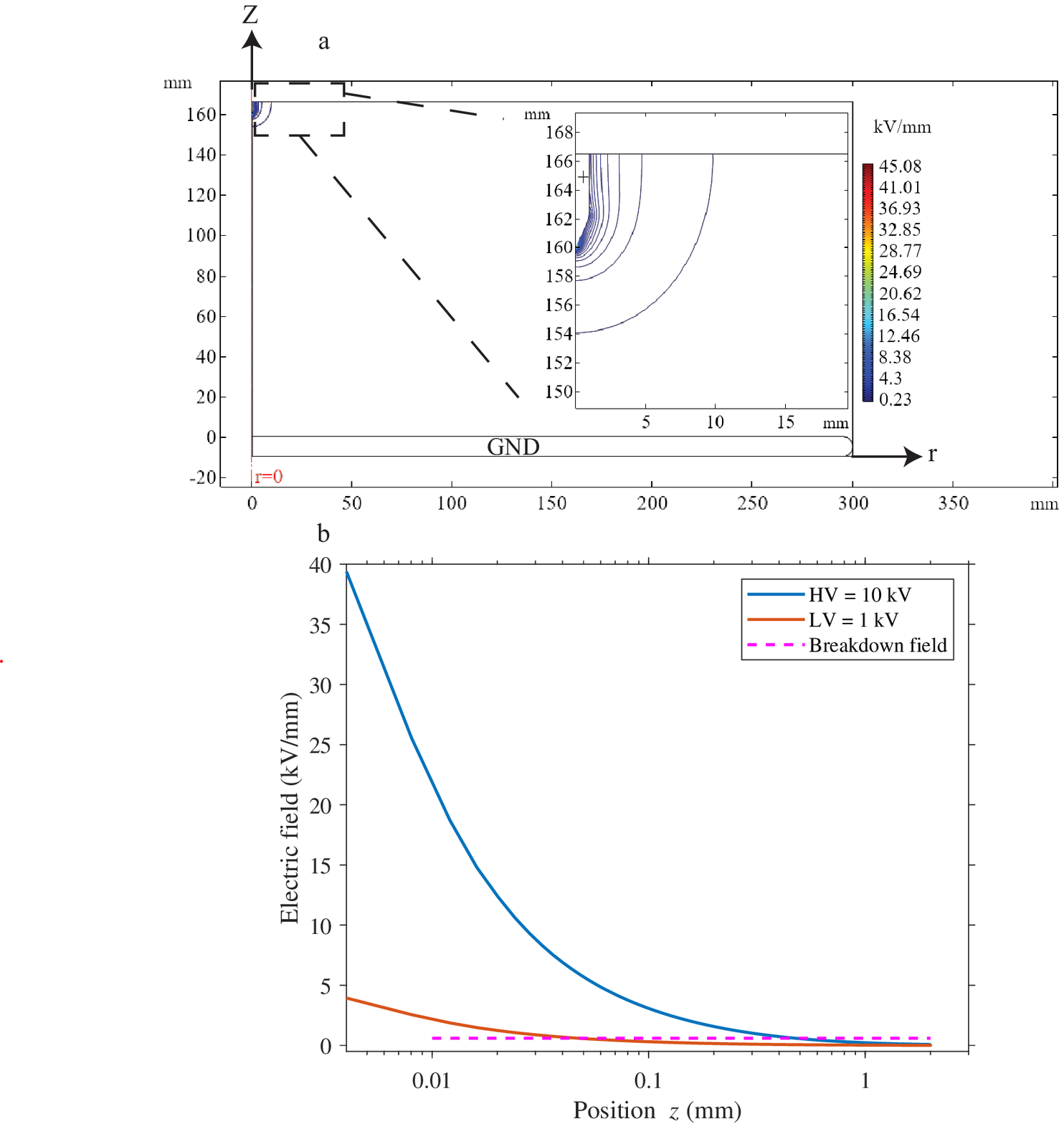}
    \caption{a) The computational domain for the electric field computation with COMSOL (included are electric field contour lines). The grounded electrode is placed at $x = 0$ and the electrode tip at $x= 160\,$mm. b) The electric field on the symmetry axis as a function of distance for the first 20\,mm from the tip for $\vhveq{}= 10\,$kV and $\vlveq{} = 1\,$kV. The breakdown electric field (dashed line) is 0.66\,kV/mm at 300\,mbar and is taken from Bagheri \textit{et al.}~\cite{bagheri2020simulation}.}
    \label{fig:E_field}
\end{figure}

\begin{figure}[ht!]
    \centering
    \includegraphics[scale=0.8]{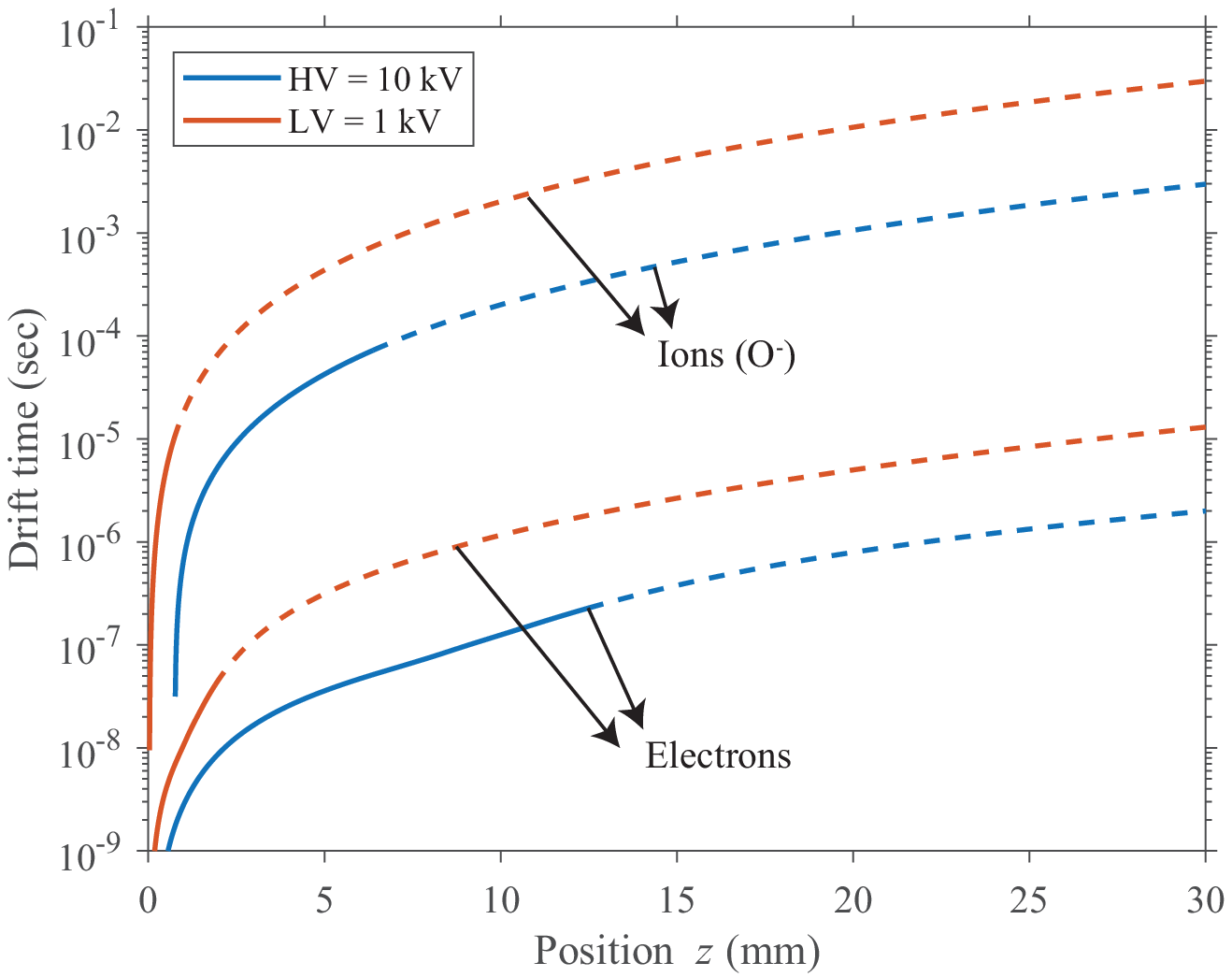}
    \caption{Drift  time calculated using equation (\ref{eq:delta_x_co2}) on  the  axis  of  symmetry  of electrons and \Omin{} ions from  a  distance $x$ to  the electrode at $x = 0$  when  a  voltage  of  $\vlveq{}= 10\,$kV or $\vlveq{} = 1\,$kV is applied across the gap. The mobilities in the unavailable electric field ranges are assumed constant, and the calculated drift times in those ranges are plotted as dashed lines.}
    \label{fig:mobility}
\end{figure}

To elaborate on this hypothesis, we have calculated the drift time of electrons and \Omin{} ions in \COtwo{} from a distance $z$ to the pointed electrode (at $z$=0) on the axis of symmetry. The drift time was calculated by 
\begin{equation}\label{eq:delta_x_co2}
    t_{\rm drift}(z) =  \int_z^0 \frac{1}{\mu(E(z')) E(z')} dz',
\end{equation}
where $\mu$, the mobility of the respective species, is derived from \cite{haefliger2018detailed} for electrons and \cite{viehland1995relating} for ions, and $E(z)$, the electric field on the axis (see figure~\ref{fig:E_field}), is calculated by a COMSOL Multiphysics simulation \cite{COMSOL}. The corresponding drift times for electrons and ions are shown in figure~\ref{fig:mobility}. It should be noted that since mobility data for electrons and ions in \COtwo{} is not available for low electric fields, the given estimations are based on constant mobility in these fields (dashed lines in figure~\ref{fig:mobility}). 

Let us now assume that we indeed have a localized \layer{} containing, amongst other species, negative ions and free electrons.
Figure~\ref{fig:mobility} enables us to investigate the initial and post-drift positions of the ion and electron distributions, \tmed{} in the baseline experiment (\vlv{} = 0) is 1.2\,\mug{}s. Figure~\ref{fig:mobility} shows that at $\vhveq{}=10\,$kV, electrons should be generated at 25\,mm from the tip to have enough time to drift and reach the tip. Hence, we can estimate the initial position of the boundary of the \layer{} to be about 25\,mm away from the tip. Note that a possible detachment time is not included in this calculation. Moreover, it is assumed that an electron should reach the tip of the electrode to initiate a discharge.

\begin{figure}[ht!]
    \centering
    \includegraphics[scale=0.6]{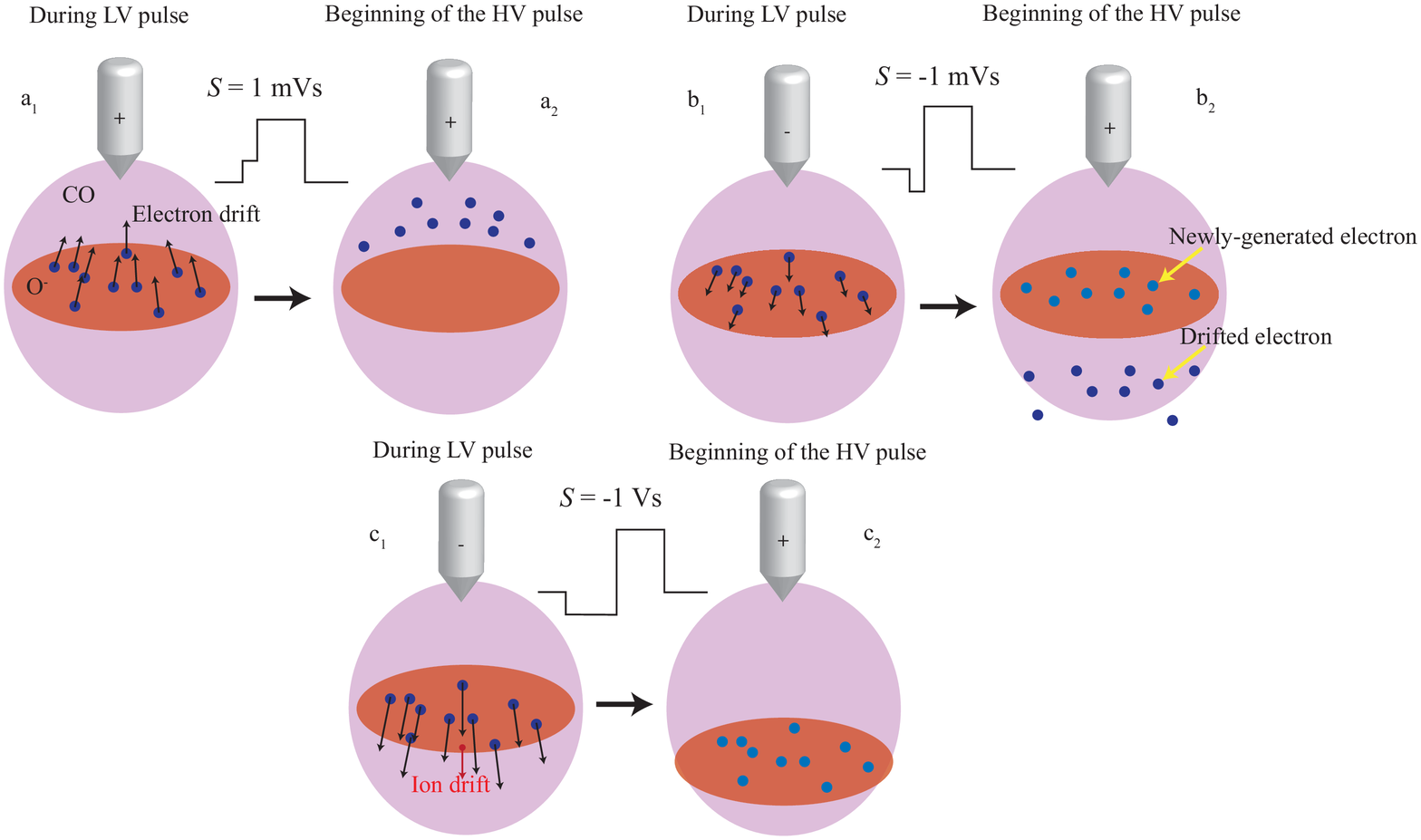}
    \caption{Visual explanation of electrons and ions drift under positive and negative LV pulses with $S=$ a) 1\,mVs, b) -1\,mVs, and c) -1\,Vs. In each figure, 1 shows the drift during an LV pulse and 2 shows the new position of electrons and ions at the beginning of the HV pulse.}
    \label{fig:drift_visual}
\end{figure}

Under a positive LV pulse (figure~\ref{fig:drift_visual}.a) with $S=1\,$mVs, the corresponding electron drift distance is around 2\,mm toward the electrode tip. During the main HV pulse, electrons with this new position require 1.01\,\mug{}s to reach the ionization zone. Therefore, we expect to observe around 182\,ns shift in the \tmed{} peak. This is relatively close to our observation from the right-hand side of figure~\ref{fig:media_time}, where the drift for every $S$ = 1\,mVs positive LV pulse was 92\,ns. According to equation~\ref{eq:delta_x_co2}, the ion drift during application of 1\,kV and 1\,\mug{}s is less than 1\,\mug{}m.

Under a negative LV pulse with $S$=1\,mVs (figure~\ref{fig:drift_visual}.b), electrons drift outwards roughly the same distance as they drift inwards when a positive LV is applied. Ions which are nearly immobile during $S$=1\,mVs, are still re-generating new electrons in the absence of the drifted electrons. This procedure also continues after the LV pulse is applied and at the beginning of the main HV pulse. The newly generated electrons are closer to the tip than their drifted cousins. Therefore, they have more chance to initiate a discharge. This supports our observation that no shift was measured in \tmed{} after an LV pulse was applied with $S$ = 1\,mVs. \Omin{} ions only start to drift noticeably under a stronger electric field. Ions under a negative LV pulse of $S$=1\,Vs drift out about 0.45\,mm from their initial position (figure~\ref{fig:drift_visual}.c). At the beginning of the main HV pulse, the electrons newly generated from the ions that are located in a new position start to drift toward the tip and imitate the discharge. According to equation~\ref{eq:delta_x_co2}, the time that electrons need to drift back 0.45\,mm is 95\,ns. This is very close to our observation in figure~\ref{fig:media_time} where the the \tmed{} shift under an LV pulse with a $S$=-1\,Vs is 92\,ns.

\begin{figure}[ht!]
    \centering
    \includegraphics[scale=0.5]{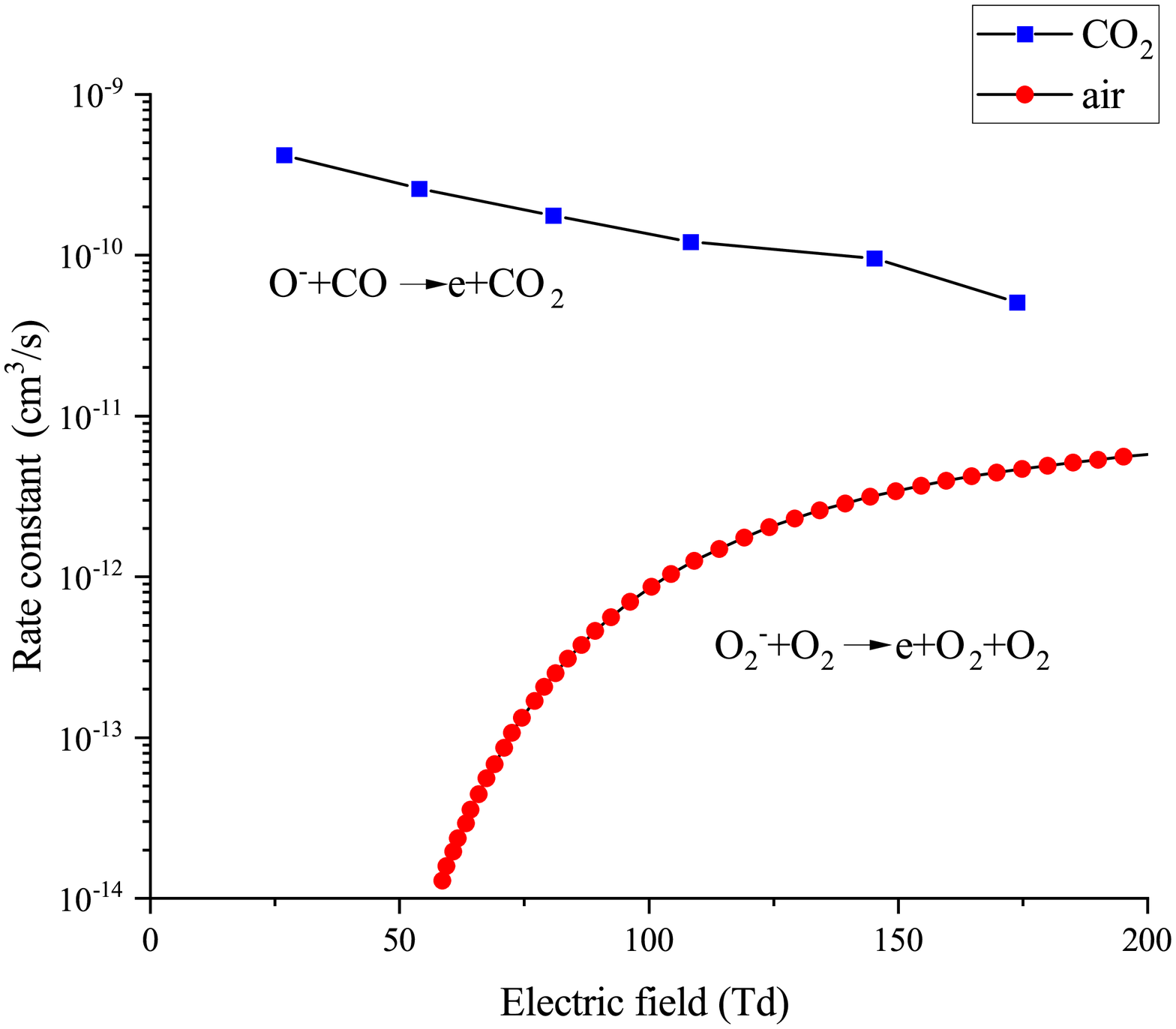}
    \caption{Rate constant of the dominant detachment processes in air and \COtwo{}. For air data was taken from \cite{pancheshnyi2013effective} and for \COtwo{} from \cite{mcfarland1973flow}.}
    \label{fig:detachment}
\end{figure}

In our previous study \cite{mirpour2020distribution}, we showed that \Otwominus{} in synthetic air was responsible for one of the peaks of the \tinc{} histogram. This peak showed a symmetric shift for positive and negative LV pulses. A possible explanation for this difference between air and \COtwo{} can be the different electron detachment mechanisms. The detachment rate of \Otwominus{} in air depends strongly on the electric field (see figure~\ref{fig:detachment}). Therefore \Otwominus{} must be in a high field region to detach an electron~\cite{pancheshnyi2013effective}.
However, for \COtwo{} the detachment rate decreases with electric field \cite{mcfarland1973flow}  (see figure~\ref{fig:detachment}). The reason is that associative detachment in \COtwo{} proceeds through autodetachment of the \cotwomin{} complex in a short period. In higher fields and energy, formation of this complex is less probable because of the shorter collision time. Thus, the associative-detachment reaction in \COtwo{} is more likely to release electrons in the low fields regions far away from the electrode tip. Therefore, during a low $S$ value LV pulse, detached electrons can move and show up their drift in \tmed{} histogram.

One question that can be raised in this section concerns the attachment of electrons during the drift. According to our hypothesis, electrons should drift about 25\,mm to reach the ionization zone. To calculate the attachment time, we exert a simple analysis using the BOLSIG+ solver \cite{BOLSIGpaper,BOLSIGprogram} with the reactions shown in table \ref{tab:reactions_co2} as input. Results show that at the electric field around 1\,Td at 2.5\,mm, where the initial electrons release from the \layer{}, the attachment time is 1.5\,\mug{}s which is slightly higher than the inception time. Therefore, this gives enough time for most of the electrons to approach the ionization zone without being attached. 

From the width of the inception time distribution in the baseline experiments and equation~(\ref{eq:delta_x_co2}) we can estimate that the thickness of the \layer{} is about 0.2\,cm. When we consider the \layer{} as disk of this thickness and with a diameter of 2.5\,cm, its volume is approximately 1\,cm$^3$. With the following equation
\begin{equation}
k_{detachment}\cdot[\textrm{CO}]\cdot[\textrm{\Omin{}}]=1/(N\cdot t),
\end{equation}
where $k_{detachment}=5.5\times10^{-10}$~\cite{wang2016effective} , we can estimate that in $t=1.5\,$\mug{}s attachment time, in order to have 1\,electron available in this volume of ($N=1/$cm$^3$), the required concentration product of [CO]$\cdot$[\Omin{}] should be of the order of magnitude 10$^{15}$\,cm$^{-6}$. When we assume that the concentrations of [CO] and [\Omin{}] are roughly equal, the density of each species should be 10$^{7}-10^{8}$\,cm$^{-3}$ in order to supply the required electrons for inception. This is a feasible number for these densities in the \layer{}.   

\begin{table} [hbt!]
    \caption{List of plasma-chemical reactions used for calculation the ionization and attachment coefficients. Cross-sections are taken from Itikawa database \cite{Itikawa_database}.}
    \centering
    \begin{tabular}{l|l}
    \hline
    Elastic collision & \elec{} + \COtwo{} $\rightarrow$ \elec{} + \COtwo{} \\
     \hline
    Ionization & \elec{} +\COtwo{} $\rightarrow$ 2\elec{} + CO$_2^+$ \\
     & \elec{} + CO $\rightarrow$ 2\elec{} + CO$^+$ \\
     \hline
    Attachment & \elec{} + \COtwo{} $\rightarrow$ CO + \Omin{}\\
     \hline
    \end{tabular}
        \label{tab:reactions_co2}
\end{table}
\subsection{Neutralization and ionization during longer LV pulse durations}
In this section we investigate higher values of $S$.
\subsubsection{Positive LV pulse}
\begin{figure}[ht!]
    \centering
    \includegraphics[scale=0.8]{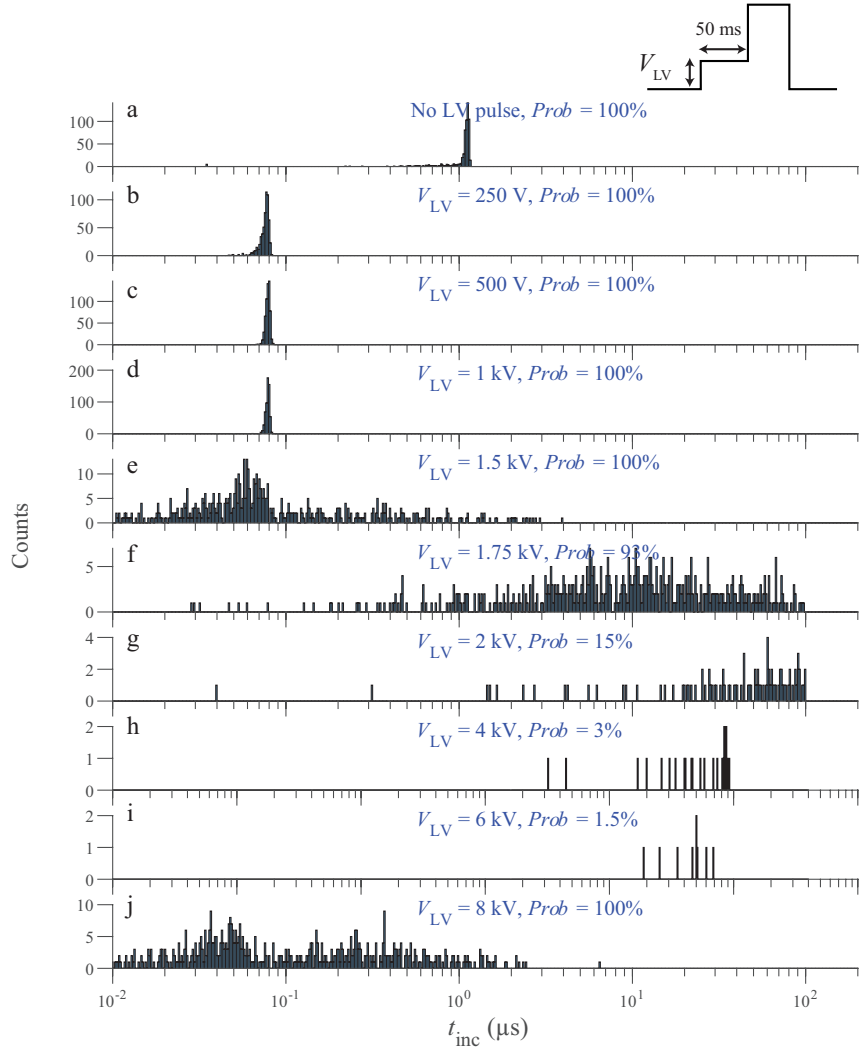}
    \caption{Histograms of discharge inception time \tinc{} for 600 discharges produced a) for no LV pulse and by applying a 50\,ms LV pulse for \vlv{} = b) 0.25, c) 0.5, d) 1, e) 1.5, f) 1.75, g) 2, h) 4, i) 6, j) 8 kV before a 10\,kV pulse of 1\,ms with a repetition frequency of 10\,Hz. $Prob$ shows the probability of inception in each configuration. }
    \label{fig:neutral}
\end{figure}

Above a certain value of $S$, \tinc{} reaches its minimum value and it is not possible to have faster inception because electrons and negative ions will have drifted all the way to the the electrode. Figure~\ref{fig:neutral} shows that applying a 250\,V and 50\,ms LV pulse was enough to initiate a discharge during the rise-time of the HV pulse. Increasing \vlv{} to 1.5\,kV was enough to drift all negative ions close to the tip. This can be verified by the mobility data in figure~\ref{fig:mobility}. Above this \vlv{}, ions reside around the electrode for a long time, where they will be slowly neutralized. Moreover, the inception probability, $Prob$, decreases with increasing \vlv{} which shows the lack of initial electrons available to initiate a streamer. We also observed this effect in our previous experiments in synthetic air \cite{mirpour2020distribution} where applying an LV pulse for a long time increased \tinc{}.

For even higher \vlv{} values, starting at 8\,kV, we observe fast inception again with a probability of 100\,\%. However, at this voltage level, the inception peak already occurs during the LV pulse, because \vlv{} itself is already high enough to lead to inception.
We observed the same effect for other values of \tlv{}. For \vlv{} below 8\,kV, the key parameter is $S$ such that for a shorter LV pulse duration, a higher LV amplitude is required to achieve the same effect. 

\subsubsection{Negative LV pulse}
\begin{figure}[ht!]
    \centering
    \includegraphics[scale=0.8]{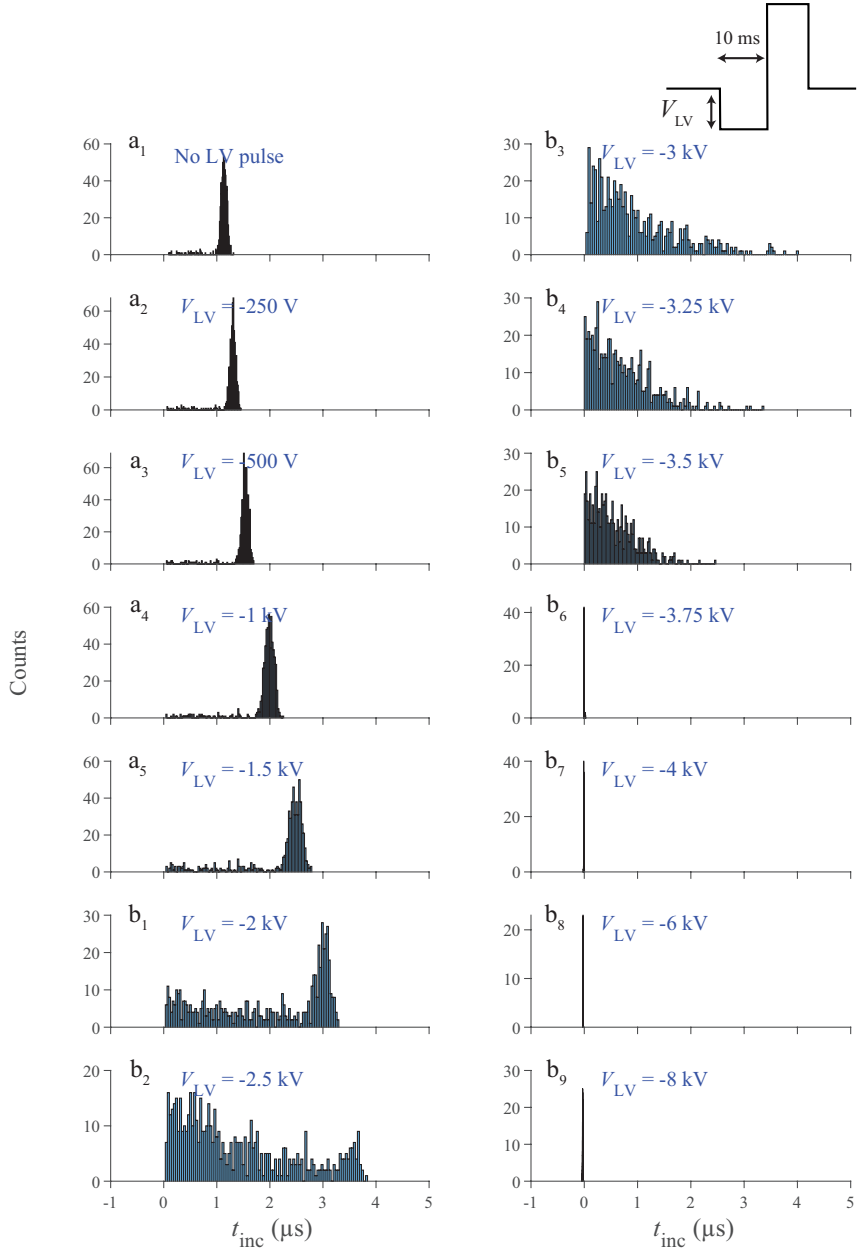}
    \caption{Histograms of discharge inception time \tinc{} for 600 discharges produced a{$_1$}) for no LV pulse and by applying a negative 10\,ms pulse for \vlv{} = a{$_2$}) -0.25, a{$_3$}) -0.5, a{$_4$}) -1, a{$_5$}) -2, (drift) b{$_1$}) -2, b{$_2$}) -2.5, b{$_3$}) -3, b{$_4$}) -3.25, b{$_5$}) -3.5, b{$_6$}) -3.75, b{$_7$}) -4 , b{$_8$}) -6, b{$_9$}) -8 kV (neutralization and ionization in LV pulse) before a 10\,kV pulse of 1\,ms with a repetition frequency of 10\,Hz.}
    \label{fig:negative_longer_lv}
\end{figure}

Similar to a long positive LV pulse, we applied a negative LV pulse with a varying \vlv{} and a fixed \tlv{} of 10\,ms (figure~\ref{fig:negative_longer_lv}). As we already discussed (left-hand side of figure~\ref{fig:media_volt}), for a lower \vlv{} we observe outward drift in which for $S  = 1\,$Vs, a 90\,ns shift in \tmed{} was observed (figure~\ref{fig:negative_longer_lv}.a${_1}$-a${_5}$). This continues until \vlv{}~=~-2\,kV where a further increase of \vlv{} leads to a peak at low \tinc{} and thereby the disappearance of the original peak in the histogram. This transition from slow \tinc{} to fast \tinc{} occurs in a small range of \vlv{}, such that at \vlv{}= -3.75\,kV \tmed{} is 30\,ns (figure~\ref{fig:negative_longer_lv}.b${_1}$-b${_6}$). Further increasing \vlv{} causes a negative \tinc{} (figure~\ref{fig:negative_longer_lv}.b${_7}$-b${_9}$). Such a negative \tinc{} indicates that the discharge initiates during the rise time of the LV pulse immediately before the main HV pulse no discharge was observed during the on-time of the LV pulse.  

\section{Summary and conclusion}
We measured inception delays \tinc{} of streamers in \COtwo{} gas to study the inception process in repetitive discharges in this gas. We applied 10\,kV high voltage pulses with a fixed pulse duration of 1\,ms and repetition frequency of 10\,Hz. This can produce some residual charges that influence \tinc{} of the next pulse. To manipulate the residual charges we applied an LV pulse with different duration \tlv{} and amplitude \vlv{} before the main HV pulses. Application of the LV pulse leads to a shift in the \tinc{} histogram peak. Three main phenomena were observed, based on the different applied $S$ (\vlv{}$\cdot$\tlv{}) values: drift, neutralization, and ionization during the LV pulse.

At low $S$ values, discharges incept faster or slower for positive and negative LV pulses respectively. However, we observed an asymmetric shift differing three order of magnitude between positive and negative LV pulses. For a positive LV pulse with $S=1\,$mVs, we observed a 92\,ns shift in \tmed{}, the median value of \tinc{}. To have a similar magnitude \tmed{} shift with a negative LV pulse, we must apply a pulse with $S=1\,$Vs. This is not consistent with our earlier observations in synthetic air. We hypothesize that this is caused by the fundamental differences between the detachment mechanism in \COtwo{} and air. \Otwominus{} ions need to drift to a high electric field region to detach an electron.
However, in the case of \COtwo{} due to the high yield of associative-detachment in low electric fields, \Omin{} ions can release electrons through reaction with CO. 
In this case, an LV pulse with a low $S$ value drifts electrons significantly and ions hardly while an LV pulse high $S$ value can also drift ions on top of the electrons.

We observed that the corresponding shift in \tmed{} due to application of LV pulse follows the mobility data and scales consistently with $S$ value. This indicates that we can neglect the effect of space charge and its effect on electric fields for low $S$ values. However we observed different phenomena for higher values of $S$.  

Applying a positive LV pulse for a longer time increases \tinc{} but decreases the probability of inception. This is likely due to neutralization of electrons and ions at the electrode. At $\vlveq{} = 8\,$kV, the LV pulse is high enough to ionize the gas and we observed streamers already during the LV pulse. This is different for a negative LV pulse where ions drift away from their initial position. For a negative LV pulse with $\vlveq{} \geq 2\,$kV, we observe formation of a fast peak (\tmed{} = 120\,ns) in the \tinc{} histogram replacing the original ion-associated peak. Increasing the voltage above 6\,kV caused discharge inception during the LV pulse rising flank.

In our calculations, electrons should drift about 25\,mm to reach the ionization zone. This seems to be a long distance since electrons can attach through several pathways such as dissociative attachment to \COtwo{} and O${_2}$, and three body attachment to O${_2}$ producing \Omin{} and \Otwominus{} \cite{aerts2015carbon}. Our simple analysis on four reactions showed that attachment time is slightly longer than inception times.